\documentclass[conference,a4paper]{IEEEtran}
\usepackage{xcolor}
\usepackage{balance}
\usepackage{xfrac}
\usepackage{float}
\usepackage{amsmath}
\ifCLASSOPTIONcompsoc
 \usepackage[caption=false,font=normalsize,labelfont=sf,textfont=sf]{subfig}
\else
 \usepackage[caption=false,font=footnotesize]{subfig}
\fi
\usepackage{url}

\begin{document}

\title{Characteristic Modes Analysis of Mutually Coupled Log-Periodic Dipole Antennas}

\author{\IEEEauthorblockN{
Georgios Kyriakou\IEEEauthorrefmark{1}\IEEEauthorrefmark{2},   
Pietro Bolli\IEEEauthorrefmark{1},   
Giuseppe Virone\IEEEauthorrefmark{3},    
}                                     

\IEEEauthorblockA{\IEEEauthorrefmark{1}
Arcetri Astrophysical Observatory, INAF, Florence, Italy, georgios.kyriakou@inaf.it, pietro.bolli@inaf.it}
\IEEEauthorblockA{\IEEEauthorrefmark{2}
Department of Physics, University of Florence, Florence, Italy}
\IEEEauthorblockA{\IEEEauthorrefmark{3}, 
CNR-IEIIT, Turin, Italy, giuseppe.virone@ieiit.cnr.it}
}

\maketitle

\begin{abstract}
Characteristic Modes Analysis (CMA) is a widely used method with recent progress in multi-antenna systems. We employ this method to characterize the mutual coupling phenomenon between two SKALA4.1 antennas, the low-frequency array elements of the future radiotelescope Square Kilometer Array (SKA-Low). The CMA accuracy is first validated at the lowest frequency range of interest with respect to a standard Method of Moments (MoM) solution by decomposing the single antenna into its characteristic modes. We then examine critical frequencies of a two-antenna system in modal decomposition, and characterize those responsible for the radiated electric field spurious spectral features owing to the mutual coupling. We connect these modes to first-order coupling of single antenna CMA modes, by using the eigenvalue data of both single- and two- antenna simulations.
\end{abstract}

\vskip0.5\baselineskip
\begin{IEEEkeywords}
characteristic modes, mutual coupling, log-periodic dipole antenna, SKA-Low
\end{IEEEkeywords}

\section{CMA review}

The description of an electromagnetic problem in terms of characteristic modes has been an objective treated in the literature dating back even to the '60s, with the first work of Garbacz \cite{garbacz1968}. Later, a rigorous representation of the problem in terms of the MoM \( Z \)-matrix was introduced by Harrington \cite{harrington1971}, which marks the start of its use in numerical problems. The method has resurfaced in the last decade, as one of its principal advantages is the physical insight it can provide to various electromagnetic problems (\cite{vogel2015}). Part of our work was heavily motivated by a recent extensive review of CMA (\cite{manteuffel2022}).

While CMA can assist in reducing the complexity of antenna design, by means of highlighting the excitation-free, inherent geometric properties of an antenna, or reducing its scope to one mode orthogonal to all others, there are some limitations to its application. Advanced eigenmode computation routines and tracking across frequency algorithms are of high computational complexity, scaling with the mode truncation order and the basis function size of the problem. Furthermore, convergence is easily achieved for far-field quantities for which the modal decomposition is complete, but near-field ones such as the reflection coefficient need ``source-current" terms to compensate for the effects of travelling-wave evanescent modes (instead of only resonant, standing-wave modes) \cite{guan2019}.

\subsection{CMA in multi-antenna systems}

CMA was initially used for single-port antennas, and in the first works of Harrington, a multi-port antenna was analysed in characteristic port currents \cite{harrington1973}. New formulations of multi-port antennas have recently been developed; the contemporary approach is to characterize the antenna in terms of the excited currents across all of the discretized geometry of a given problem, while port quantities are used only in equivalent network representations. 

Within these works, there are a number of methods for processing the mutual coupling between two antennas. Raines \cite{raines2011} has presented the treatment of the system by block-partitioning the \( Z \)-matrix in 4 blocks (2 diagonal self-terms and 2 off-diagonal that describe the coupling), and applying the standard theory to its Schur complement, which characterizes the active antenna in presence of the passive one. Application of these techniques by element loading has been considered in \cite{wu2016,ludick2019}. Ghosal \cite{ghosal2022} and Schab \cite{schab2016} have used the theory of coupled modes to directly partition the problem into uncoupled and coupled linear ``subspaces'', aiming to understand the phenomenon from a more algebraic point of view. 


\subsection{CMA in log-periodic antennas}

Log-periodic dipole antennas can present mutual coupling phenomena in array configurations, especially when the excited polarization is aligned with the translation axis, and the antennas are close to each other. This is directly related to coupling of pairs of dipoles of distinct antennas, which can well be described with standard methods, such as the Method of Moments (MoM). For an array of dipoles of arbitrary lengths, recent work examining their modes has connected the analytic methods with the CMA formulation \cite{lonsky2018}. This work can directly be extended to a single log-periodic antenna which is loaded by its boom; though arrays of such antennas have not appeared in the literature. We will attempt here to analyse a specific log-periodic design, as a single element and in a configuration of a pair of antennas. 

The antenna of interest is a SKALA4.1 \cite{bolli2020}, an active, dual-polarized log-periodic antenna that has been selected as the array element of SKA-Low \cite{dewdney2009}. While the array of 256 elements is in a pseudo-random layout, there are instances where two antennas almost align, enhancing the effect of mutual coupling. In a previous work, critical frequencies where this happens were analysed, as well as possible redistribution of the antennas in a new layout to mitigate the Embedded Element Pattern gain disturbance reported in \cite{bolli2022}. It is though of increased interest to understand the coupling mechanism of these antennas in greater detail, and identify any possible solutions to fully compensate for it. 

\section{Single antenna analysis \label{sec:single_antenna}}
\begin{figure}
    \centering
    \includegraphics[width=\columnwidth]{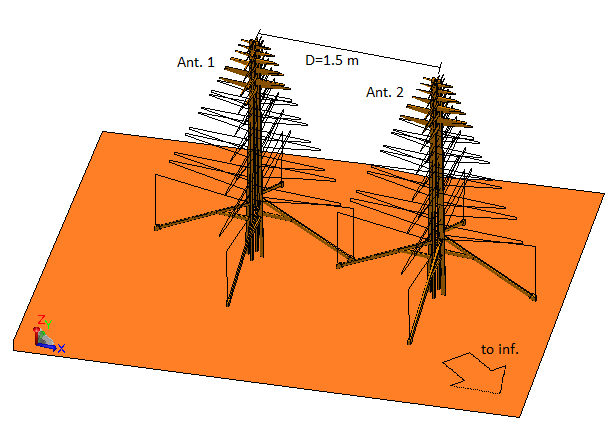}
    \caption{Two SKALA4.1 antennas placed at a distance of \( D=1.5\ m \) (center-to-center) on the \(x\)-axis. Ant.1 is the active antenna (X-polarized excitation) and Ant.2 is the passive antenna (\( 50\ \Omega \) termination). Both are placed on an infinite PEC ground plane.}
    \label{fig:capture}
\end{figure}

We will focus our analysis in the frequency range 50-100 MHz. This is a critical region for the scientific objectives of SKA-Low, since the Cosmic Dawn science is expected to be conducted there \cite{labate2017}. Some specific features of the SKALA4.1 antenna are the use of 19 triangular dipoles per polarization along with a bow-tie one, and its connection to a ground plane (here considered infinite) by a rectangular boom. The geometry of two antennas, as will be examined in Sec.~\ref{sec:two_antennas}, is seen in Fig.~\ref{fig:capture}.

\begin{figure}
    \centering
    \subfloat[]{\includegraphics[width=\columnwidth]{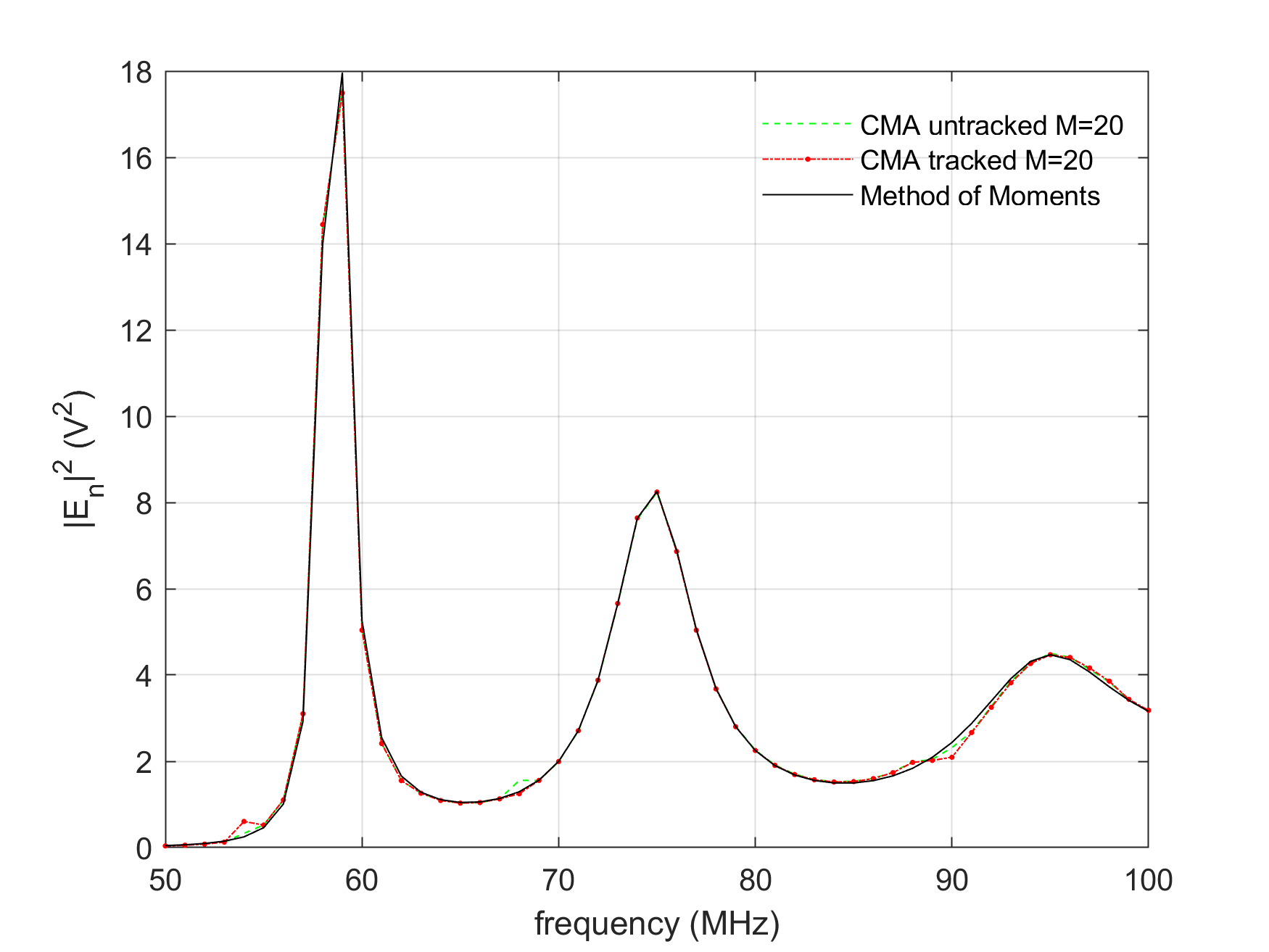}\label{fig:left1}}
    \hfil
    \subfloat[]{\includegraphics[width=\columnwidth]{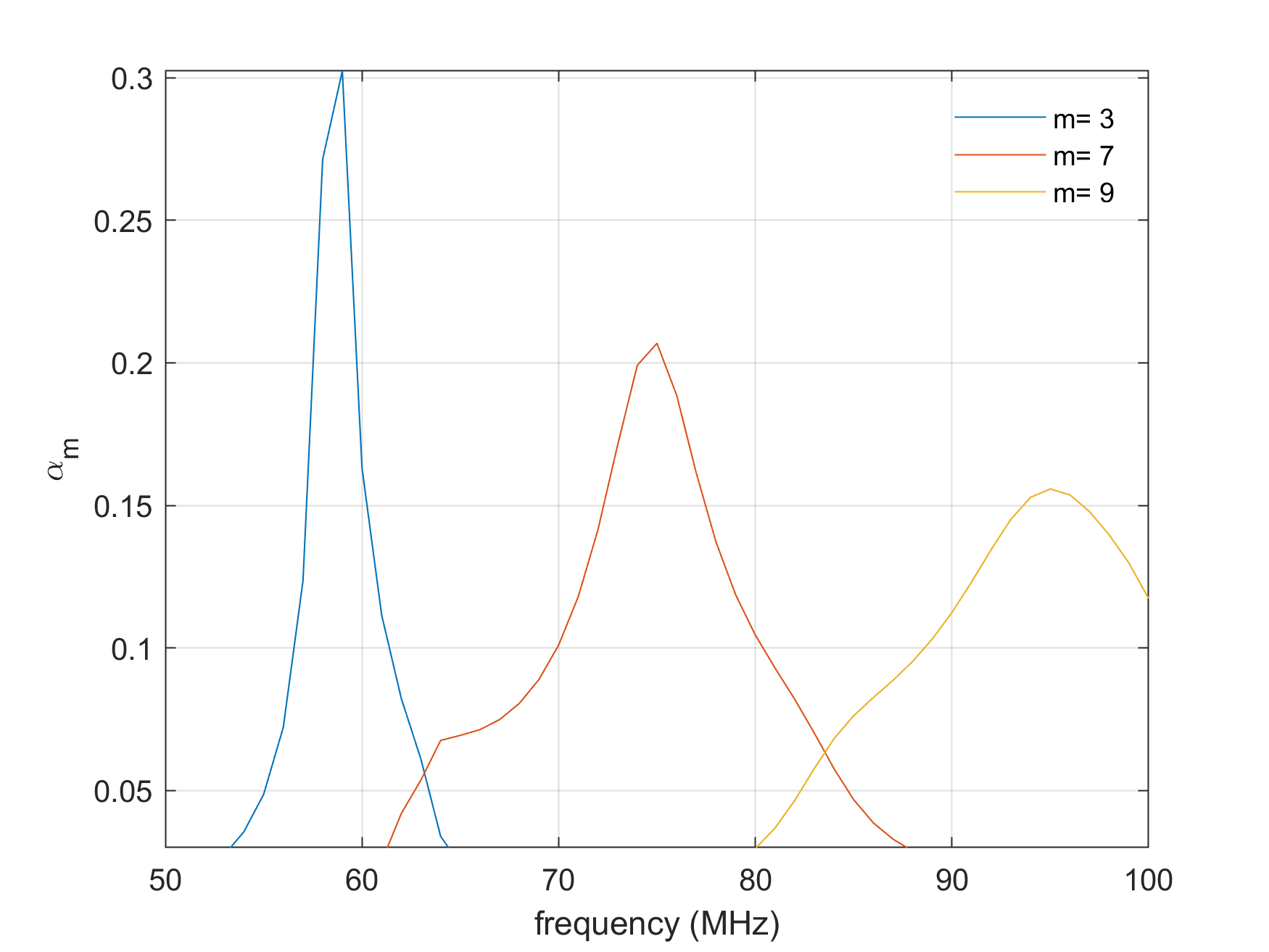}\label{fig:right1}}
    \caption{Up: Square modulus of normalized electric far-field  values at zenith across frequency, using MoM and CMA with \( M=20 \) modes. Even without the tracking routine, the CMA solution converges to that of the MoM. Down: Tracking of the normalized excitation coefficient \( \alpha_m \) across frequency of the three most significant modes, two of them (\( m=3,\ m=9 \)) directly associated to the dipole resonances at the respective peak of their \( \alpha_m \).}
    \label{fig:single_antenna}
\end{figure}

For our electromagnetic simulations, we use FEKO\footnote{https://altairhyperworks.com/feko/}, which implements CMA (thoroughly tested in \cite{capek2017}) as well the MoM. The size of a single antenna (\( \approx 9000 \) DoF) is challenging for experimentation with multiple solvers, so this initial work will be focused only on FEKO. We excite the SKALA4.1 antenna at its X-polarized dipoles with a constant voltage of \( V_s=1\ V \), and use a number of \( M=20 \) modes for the approximation of the solution, heuristically chosen the same as the number of dipoles per polarization of the antenna. The frequency step is 1 MHz. In Fig.~\ref{fig:left1}, we present a comparison of the squared modulus of the normalized (with respect to radial distance \( r \)) electric far-field values, \( |E_n|^2=r^2\left(|E_\theta|^2+|E_\phi|^2\right) \), at zenith, computed using both the MoM and CMA solvers. An important step in CMA is the eigenmode tracking across frequency, that is necessary to avoid confusion of the physical properties of each mode. In other words, it is the continuity of the eigenmode as a function of frequency that the eigenvalue calculator cannot standalone ensure. We present both the tracked and the untracked solution, to show that this number of modes can succeed in converging to the MoM without tracking. It can be seen that both solutions generally converge, with a number of small anomalies (see green curve in Fig.~\ref{fig:left1}) at 54, 68 and 90 MHz. The regions around 59, 75 and 95 MHz contain the peaks of the antenna radiation in this range, also observed as \( |S_{11}| \) local minima in \cite{bolli2020}.

Such peaks are directly related to dipole modes, namely the radiation resonances of the bow-tie dipole and dipole 19, the next one on the \( +z \) axis, at frequencies 59 and 95 MHz respectively. We next present the tracking across frequency of the three most significant modes in terms of their normalized excitation coefficient
\begin{equation}
    \alpha_m=\frac{{J}_m(\mathbf{r}_s)\cdot V_s}{|1+j\lambda_m| (\vec{J}_{m}^T\Re\{Z\}\vec{J}_{m})}     
    \label{eq:alpha}
\end{equation} 

where \( {J}_m(\mathbf{r}_s)\cdot V_s \) is the Modal Excitation Coefficient (MEC) as the inner product between the current eigenvector at the source location \( \mathbf{r}_s \) and the voltage excitation, and \( \lambda_m \) the respective eigenvalue. In Fig.~\ref{fig:right1}, we have identified modes \( m=3,\ m=7\ \textrm{and}\ m=9 \) (as labelled by FEKO) as those responsible for the radiation peaks reported. Some tracking issues have appeared in the examined frequency range for less significant modes, a motive for which we lower-thresholded the \(y\)-axis of Fig.~\ref{fig:right1} at 0.03; they will though not affect our analysis as long as a constant mode ordering has been chosen for the frequencies across which the solution is properly modally tracked. Another noteworthy characteristic of the tracked modes are the crossings between them, meaning that at certain frequencies, some eigenvalues appear with multiplicity greater than one. According to \cite{schab2017}, this is an allowed feature of antennas belonging to certain symmetry groups (SKALA4.1 is symmetric with respect to a \( 180^{\circ} \) rotation), while it can also serve as an evaluation of the tracking algorithm. It should also be clarified that if we do not include MEC in \( \alpha_m \) we get the Modal Significance (MS), which is a characteristic of the impedance matrix independent of the excitation placement. We do not show this quantity but it has been verified to have the same trend.

\section{Pair of antennas\label{sec:two_antennas}}

We now focus our analysis in a configuration of two SKALA4.1 antennas in close proximity. We place the second antenna at a distance of \( D=1.5\ m \) on the \(x\)-axis, since we want to examine mutual coupling phenomena at the configuration corresponding to their maximum effect, which is when they perfectly align in their E-plane. It should be taken into account that, in this way, we approximately create another geometric symmetry on the problem (that of mirroring with respect to the \( Oyz \)-plane), which will certainly give rise to characteristic mode traces almost crossing, that increase the difficulty of tracking \cite{schab2017}. 

\begin{figure}
    \centering
    \subfloat[]{\includegraphics[width=\columnwidth]{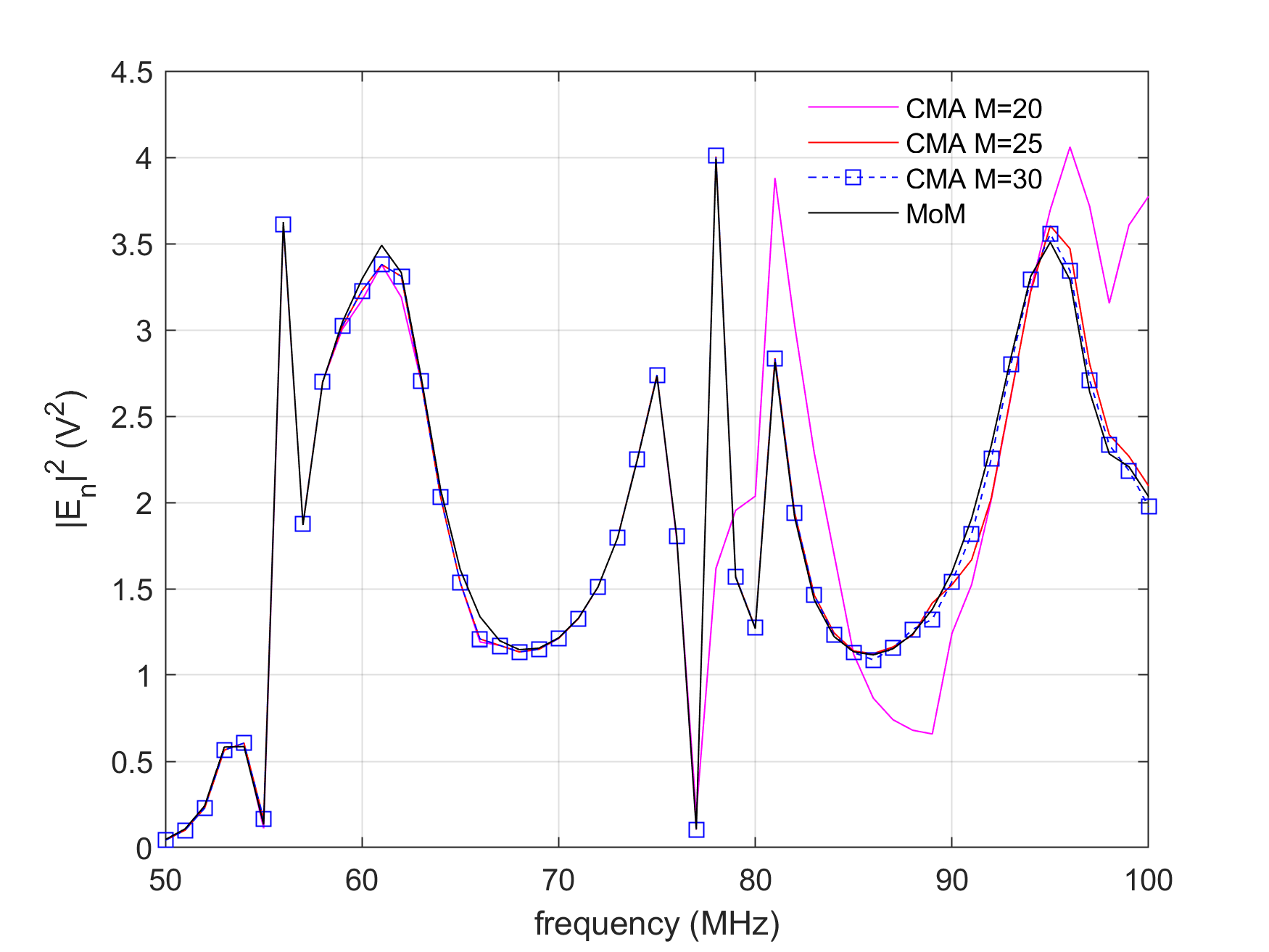}\label{fig:left2}}
    \hfil
    \subfloat[]{\includegraphics[width=\columnwidth]{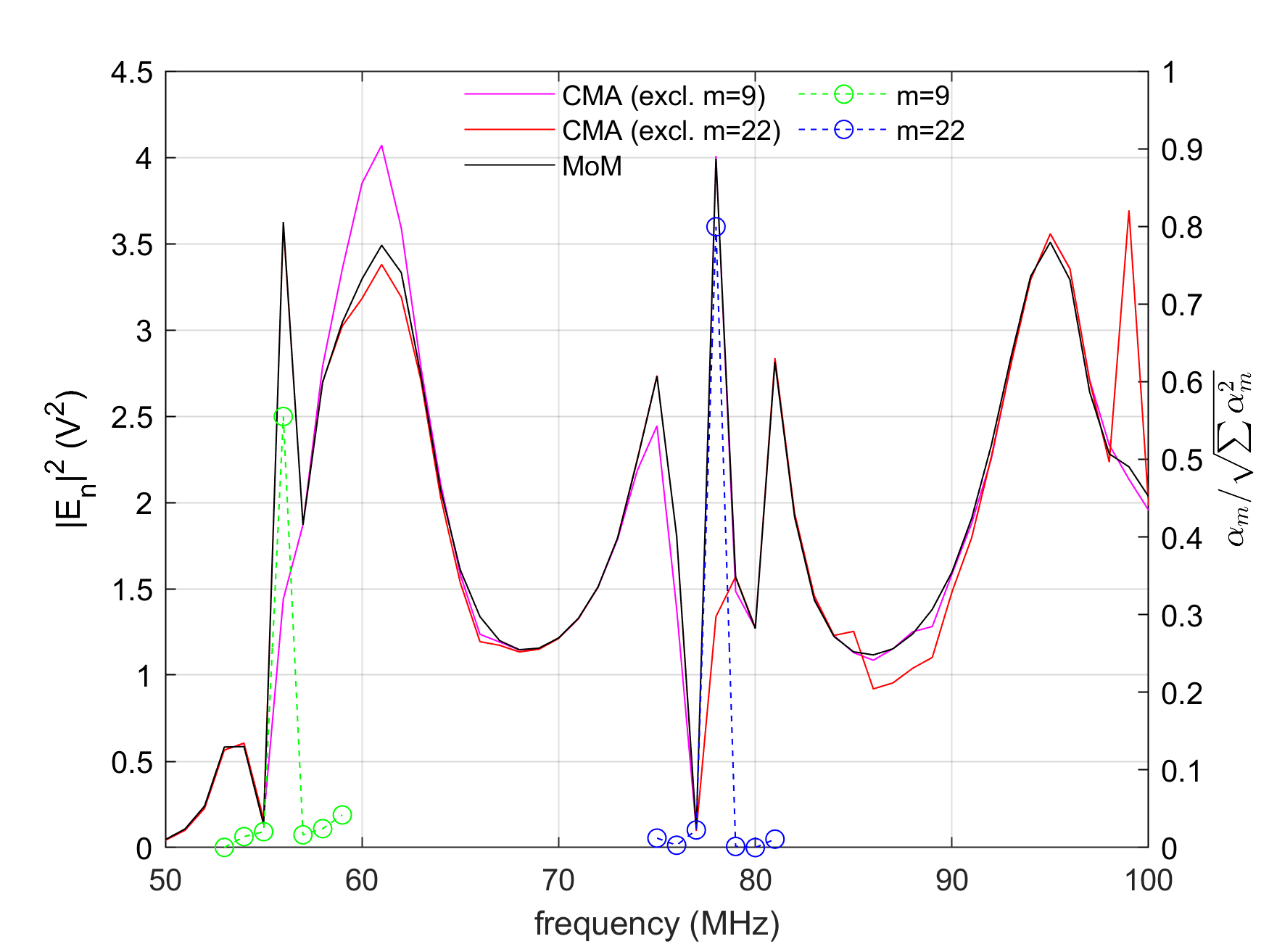}\label{fig:right2}}
    \caption{Up: Square modulus of normalized electric far-field  values at zenith across frequency, using MoM and CMA with \( M=20,\ 25 \) and \( M=30 \), Down: left axis and legend column correspond to the same electric field values using MoM and CMA but excluding modes \( m=9,\ m=22 \) respectively, while the right shows the corresponding \( \alpha_m \) properly normalized to \( \leq 1 \).}
    \label{fig:pair_antennas}
\end{figure}

The simulation procedure discussed in Sec.~\ref{sec:single_antenna} has been repeated on the antenna pair using a larger number of modes to achieve convergence. We first present in Fig.~\ref{fig:left2} the same zenith far-field \( |E_n|^2 \) values, starting by \( M=20 \) and converging at \( M=30 \) to the MoM solution. We note that the values of \( |E_n|^2 \) are lower than that of Fig.~\ref{fig:left1}, meaning that the radiated power is also lower. As has been shown though in \cite{bolli2022}, the gain remains relatively constant with respect to that of the single antenna, at the frequencies where the mutual coupling is not significant. 

We observe that the 3 distinct radiation peaks, a characteristic inherited from the active antenna, are now disturbed; due to the poor tracking though we were not able to directly associate these features with their distinct modes. It must be highlighted that the CMA solutions could not be sufficiently tracked even at a post-processing stage. The increased complexity and size of the problem result in FEKO's tracking routine further lacking in accuracy \cite{capek2017}. We also identify the mutual coupling features, that discontinuously alter the smooth response, at frequencies 56 and 78 MHz, as reported in \cite{bolli2022}. Since these are narrow-band, we have tentatively tried to track the solutions as shown in Fig.~\ref{fig:right2} with circle-dashed lines and values on the right axis, which correspond to the \( \alpha_m/\sqrt{\sum\alpha_m^2} \) (normalized to \( \leq 1 \), proving that those are the dominant modes in the respective narrow bands). To enhance the understanding of this result, we also show the reconstructed CMA solution by respective omission of these modes. The resulting \( |E_n|^2 \) curves (labelled ``excl. m=9'', ``excl. m=22'') are presented with values on the left axis of Fig.~\ref{fig:right2}, resulting in continuity of the curve at each narrow frequency band by omitting the respective mode.

We will now present the connection between modes of the single antenna and modes of the system of two antennas, using the theory of coupled characteristic modes as outlined in \cite{ghosal2022}. We are going to assume the first-order approximation by considering the \( m \)-th coupled characteristic mode eigencurrent of interest as:

\begin{equation}
    \vec{J}_m=\begin{bmatrix}
    \vec{J}_{m_1} \\ c\vec{J}_{m_2}
    \end{bmatrix}
\end{equation}

where \( \vec{J}_{m_1},\ \vec{J}_{m_2} \) are the interacting \(m_1\)-th, \(m_2\)-th eigencurrents of each single antenna, respectively, and \( c \) is a scalar coupling constant. Since we are interested in modes \( m=9,\ m=22 \) of the coupled system, we would need to find the respective \( m_1,\ m_2 \) for each case. It has to be emphasized that by using the coupling coefficient it is implicitly assumed that Ant.1 is not influenced by Ant.2; it rather exerts a perturbation on it \cite{ghosal2020}. This way, \( c\alpha_{m_1} \) could be thought of as a new factor \( \alpha_{m_2,eff} \), quantifying the first order effect as an effective normalized excitation coefficient of mode \( m_2 \) on Ant.2 by means of the applied excitation of mode \( m_1 \) of Ant.1.

Examining the eigenvalues of both the single and the two-antenna system at frequencies 56 and 78 MHz, where modes \( m=9 \) and \( m=22 \) respectively are dominant causing a mutual coupling glitch, has resulted in the identification of the dominant modes \( m_1 \) of Ant.1 perturbed in the presence of Ant.2. The single-antenna eigenvalue \( \lambda_{m_1} \) along with the two-antenna system eigenvalue \( \lambda_m \) and their percentage relative difference \( |\lambda_{m_1}-\lambda_m|/|\lambda_{m_1}| \) for both the examined frequencies are listed in Tab.~\ref{tab:coupled_modes}. At 56 MHz, the relative difference seems to be a quite larger perturbation of the single-antenna eigenvalue, while a more reasonable such value is calculated at 78 MHz. In retrospect, \( \lambda_{m_2} \), an ``excited'' mode that has not been identified, could be non-dominant for the single antenna, but still have a large \( a_{m_2,eff} \) in the antenna pair and influence the perturbation of the exciting mode. This aspect is already under further study and validation by applying the coupled mode theory in a more rigorous way to fully characterize the first-order coupling.

\begin{table}
\caption{Eigenvalues of single-antenna mode, two-antenna system mode and relative difference}
\label{tab:coupled_modes}
\centering
\begin{tabular}{|c||c|c|c|c|}
\hline
f (MHz) & \( \lambda_{m_1} \) & \( \lambda_m \) & \( \left|\frac{\lambda_{m_1}-\lambda_m}{\lambda_{m_1}}\right|\cdot100\% \)\\
\hline
56 & -0.489 & -0.608 & 24.3\% \\
\hline
78 & 1.854 & 1.849 & 0.2\% \\
\hline
\end{tabular}
\end{table}
\section{Conclusions}
We have used the Characteristic Modes analysis employing a log-periodic dipole antenna, the SKALA4.1, in a single and two-antenna configuration. The size of the problem is challenging with respect to the interpretation of its results. CMA is proven to converge to the MoM solution under selection of a sufficient number of modes. Tracking issues prevent us from fully understanding the behaviour of each mode across the examined frequency range, but narrow-band mutual coupling glitches are sufficiently tracked and localized on a certain modal component. The theory of coupled characteristic modes is used to connect the mutual coupling radiation modes identified in the two-antenna system with a dominant mode of the active antenna viewed as single in each case. 

\balance


\begin{thebibliography}{1}

\bibitem{garbacz1968}
R.~J.~Garbacz, ``A generalized expansion for radiated and
scattered fields,'' Ph.D. dissertation, Ohio State University, Columbus, 1968. 

\bibitem{harrington1971}
R.~Harrington and J.~Mautz, ``Theory of characteristic modes for conducting bodies,'' in \emph{IEEE Transactions on Antennas and Propagation}, vol. 19, no. 5, pp. 622-628, September 1971, doi: 10.1109/TAP.1971.1139999.

\bibitem{vogel2015}
M.~Vogel, G.~Gampala, D.~Ludick, U.~Jakobus and C.~J.~Reddy, ``Characteristic Mode Analysis: Putting Physics back into Simulation,'' in \emph{IEEE Antennas and Propagation Magazine}, vol. 57, no. 2, pp. 307-317, April 2015, doi: 10.1109/MAP.2015.2414670.

\bibitem{manteuffel2022}
D.~Manteuffel, F.~H.~Lin, T.~Li, N.~Peitzmeier and Z.~N.~Chen, ``Characteristic Mode-Inspired Advanced Multiple Antennas: Intuitive insight into element-, interelement-, and array levels of compact large arrays and metantennas,'' in \emph{IEEE Antennas and Propagation Magazine}, vol. 64, no. 2, pp. 49-57, April 2022, doi: 10.1109/MAP.2022.3145714.

\bibitem{guan2019}
L.~Guan, Z.~He, D.~Ding and R.~Chen, ``Efficient Characteristic Mode Analysis for Radiation Problems of Antenna Arrays,'' in \emph{IEEE Transactions on Antennas and Propagation}, vol. 67, no. 1, pp. 199-206, Jan. 2019, doi: 10.1109/TAP.2018.2876705.

\bibitem{harrington1973}
J.~Mautz, R.~Harrington, ``Modal analysis of loaded N-port scatterers,'' in \emph{IEEE Transactions on Antennas and Propagation}, vol. 21, no. 2, pp. 188-199, March 1973, doi: 10.1109/TAP.1973.1140431.

\bibitem{raines2011}
B.~D.~Raines, ``Systematic Design of Multiple Antenna Systems using Characteristic Modes,'' Ph.D dissertation, Ohio State University, Columbus, 2011.

\bibitem{wu2016}
Q.~Wu, W.~Su, Z.~Li and D.~Su, ``Reduction in Out-of-Band Antenna Coupling Using Characteristic Mode Analysis'' in \emph{IEEE Transactions on Antennas and Propagation}, vol. 64, no. 7, pp. 2732-2742, July 2016, doi: 10.1109/TAP.2016.2522459

\bibitem{ludick2019}
D.~J.~Ludick, ``Applying the Theory of Characteristic Modes to the Analysis of Finite Antenna Array elements and Ground Planes of Finite Sizes,'' in \emph{2019 13th European Conference on Antennas and Propagation (EuCAP)}, 2019, pp. 1-4.

\bibitem{ghosal2022}
S.~Ghosal, R.~Sinha, A.~De and A.~Chakrabarty, ``Characteristic Mode Analysis of Mutual Coupling,'' in \emph{IEEE Transactions on Antennas and Propagation}, vol. 70, no. 2, pp. 1008-1019, Feb. 2022, doi: 10.1109/TAP.2021.3119117.

\bibitem{schab2016}
K.~R.~Schab, J.~M.~Outwater, M.~W.~Young and J.~T.~Bernhard, ``Eigenvalue Crossing Avoidance in Characteristic Modes,'' in \emph{IEEE Transactions on Antennas and Propagation}, vol. 64, no. 7, pp. 2617-2627, July 2016, doi: 10.1109/TAP.2016.2550098.



\bibitem{lonsky2018}
T.~Lonsky, P.~Hazdra and J.~Kracek, ``Characteristic Modes of Dipole Arrays,'' in \emph{IEEE Antennas and Wireless Propagation Letters}, vol. 17, no. 6, pp. 998-1001, June 2018, doi: 10.1109/LAWP.2018.2828986.

\bibitem{bolli2020}
P. Bolli \textsl{et al.}, ``Test-Driven Design of an Active Dual-Polarized Log-Periodic Antenna for the Square Kilometre Array,'' in \emph{IEEE Open Journal of Antennas and Propagation}, vol. 1, pp. 253-263, 2020, doi: 10.1109/OJAP.2020.2999109.

\bibitem{dewdney2009}
P.~E.~Dewdney, P.~J.~Hall, R.~T.~Schilizzi and T.~J.~L.~W.~Lazio, ``The Square Kilometre Array,'' in \emph{Proceedings of the IEEE}, vol. 97, no. 8, pp. 1482-1496, Aug. 2009, doi: 10.1109/JPROC.2009.2021005.

\bibitem{bolli2022}
P.~Bolli, M.~Bercigli, P.~Di Ninni, L.~Mezzadrelli and G.~Virone, ``Impact of mutual coupling between SKALA4.1 antennas to the spectral smoothness response,'' in \emph{J. Astron. Telesc. Instrum. Syst.}, 8(1), pp. 11-23, 25 Feb. 2022, doi: 10.1117/1.JATIS.8.1.011023

\bibitem{labate2017}
M.~G.~Labate, R.~Braun, P.~Dewdney, M.~Waterson and J.~Wagg, ``SKA1-LOW: Design and scientific objectives,;; in \emph{XXXIInd General Assembly and Scientific Symposium of the International Union of Radio Science (URSI GASS)}, 2017, pp. 1-4, doi: 10.23919/URSIGASS.2017.8105424.

\bibitem{capek2017}
M.~Capek, V.~Losenicky, L.~Jelinek and M.~Gustafsson, ``Validating the Characteristic Modes Solvers,'' in \emph{IEEE Transactions on Antennas and Propagation}, vol. 65, no. 8, pp. 4134-4145, Aug. 2017, doi: 10.1109/TAP.2017.2708094.

\bibitem{schab2017}
K.~R.~Schab and J.~T.~Bernhard, ``A Group Theory Rule for Predicting Eigenvalue Crossings in Characteristic Mode Analyses,'' in \emph{IEEE Antennas and Wireless Propagation Letters}, vol. 16, pp. 944-947, 2017, doi: 10.1109/LAWP.2016.2615041.

\bibitem{ghosal2020}
S.~Ghosal, R.~Sinha, A.~De, A.~Chakrabarty and H.~Son, ``Theory of Coupled Characteristic Modes,'' in \emph{IEEE Transactions on Antennas and Propagation}, vol. 68, no. 6, pp. 4677-4687, June 2020, doi: 10.1109/TAP.2020.2969852.

\end{thebibliography}
\end{document}